\documentclass[epj]{svjour}

\usepackage{graphics}
\usepackage{color}
\usepackage{amsmath,amssymb}
\usepackage{times}
\usepackage{ae,aeguill}
\onecolumn

\begin{document}
\title{Modulational instabilities in Josephson oscillations of 
elongated coupled condensates}
 \author{Isabelle Bouchoule} \institute{
 Laboratoire Charles Fabry de l'Institut d'Optique, UMR 8501 du CNRS, 
 91403 Orsay, France
  }

\abstract{
 We study the Josephson oscillations of two coupled elongated condensates. 
Linearized calculations  show that the oscillating mode uniform over 
the length of the condensates (uniform Josephson mode)
is unstable : modes of non zero longitudinal momentum grow exponentially.
In the limit of strong atom interactions, 
we give  scaling laws for the instability time constant
and unstable wave vectors. 
Beyond the linearized approach, numerical calculations show 
a damped recurrence behavior :
the energy in the Josephson mode presents damped oscillations. 
 Finally, we derive 
conditions on the confinement 
of the condensates to prevent instabilities.
}

\authorrunning{I. Bouchoule}
\titlerunning{Modulational instabilities in Josephson oscillations}

\maketitle

\section{Introduction}
 Josephson oscillations arise between two Bose-Einstein condensates
coupled by  tunneling effect. 
 They have been observed in superfluid 
Helium\cite{josephsonHe} and in superconductors\cite{LivreJosephsonsupra}
and have recently been achieved in dilute atomic BEC in a double
well potential\cite{Josephson-Oberthaler}.
 The physics of two coupled condensates has been extensively studied
in a two modes model, where only two single particle modes are 
involved\cite{Legett2001,Sols1999}.
 For atoms interacting in each well through a two-body interaction,
different regimes are reached depending on the ratio between the 
 tunneling strength to the  interaction energy of 
atoms in each well\cite{Smerzi97_Josephson,Legett2001}. 
For small  interaction energy,  one expects
to observe Rabi oscillations. For large  interaction energy
one enters the Josephson regime. 
 In this regime, oscillations around equilibrium configuration 
have a reduced amplitude 
in atom number and their frequency depends on the mean field energy. 
 Finally, for very large interaction energy, quantum fluctuations 
are no longer negligible : the system is  in the so called Fock regime 
and oscillations of atoms between the wells do not occur any more.
 In this paper, we assume this regime is not reached. Oscillations
between the two wells, both in the Rabi and in the Josephson regime,
are then well described by a mean field approach.

 Atom chips\cite{Folmanrevue} are probably good candidates 
to realize Josephson oscillations of Bose-Einstein Condensates 
as they enable the realization of micro-traps with strong confinement
and  flexible geometries.
A possible configuration to realize a tunnel 
coupling between BEC
on an atom-chip is  proposed in \cite{SchummJosephson}.
 In this proposal, 
the two  condensates are very elongated and 
are coupled all along their longitudinal extension. 
 With such an elongated geometry, 
both the Rabi  and the Josephson regime could be accessed. 
 However, in this case, tunnel coupling may be larger than the longitudinal
frequency and the two modes model a priori breaks down.
 In this paper, we are interested in the stability 
of the uniform Josephson mode where all the atoms oscillate between the two
wells independently of their longitudinal position.
 In the absence of 
interaction between atoms and if the transverse and longitudinal trapping 
potentials are separable, the longitudinal and transverse degree of freedom
are decoupled and one expects to observe stable Rabi oscillations between the 
condensates. On the other hand interactions between atoms introduce non linearities that
may couple the two motions.
 For a homogeneous situation such as atoms trapped 
in a box-like potential,  uniform Josephson oscillations are
a solution of the mean field evolution equations and are a priori 
possible, even in presence of interactions between atoms.
 However, the non linearities introduced by interactions between atoms
may cause instability of this  uniform Josephson mode. 
 Similar modulational instabilities  appear 
in many situations of nonlinear 
physics such as water waves propagation\cite{BenjaminandFeir}
or light propagation in a non 
linear fiber\cite{Tai86}.
 In the context of Bose Einstein condensates, they have been observed
in presence of a periodic potential, at positions 
in the Brillouin zone where the effective mass 
is negative\cite{Fallani04,Bia01,Konotop02}. 
 In our case  a modulational instability would cause
uniform Josephson oscillations to
decay into modes of non vanishing longitudinal 
momentum. The goal of this paper is to investigate those instabilities. 

 We assume that all the relevant frequencies (interaction energy and 
tunnel coupling) are much smaller than the transverse oscillation
frequencies in each well so that  we can consider only a one dimensional 
problem.
Thus, the system we consider is described by the Hamiltonian
\begin{equation}
 \begin{array}{lll}
 H &=& \int dz\left\{\frac{-\hbar^2}{2m}
 \left[\psi^{\dag}_1(z)\frac{\partial^2}{\partial z^2}\psi_1(z) +
 \psi^{\dag}_2(z)\frac{\partial^2}{\partial
 z^2}\psi_2(z)\right]\right.
\\
 && + U(z)\left[\psi^{\dag}_1(z)\psi_1(z) +
 \psi^{\dag}_2(z)\psi_2(z)\right]
\\
 && +
 \frac{g}{2}\left[\psi^{\dag}_1(z)\psi^{\dag}_1(z)\psi_1(z)\psi_1(z)
 + \psi^{\dag}_2(z)\psi^{\dag}_2(z)\psi_2(z)\psi_2(z)\right]
\\
&&\left.- \gamma\left[\psi^{\dag}_1(z)\psi_2(z) +
 \psi^{\dag}_2(z)\psi_1(z)\right]\right\},
 \end{array}
\label{ham}
\end{equation}
where $g$ is the one-dimensional coupling constant and 
$U(z)$ is the longitudinal trapping potential. 
For a harmonic transverse confinement for which $\omega_{\perp} \ll \hbar^2/(ma^2)$,
we have $g=2\hbar\omega_{\perp}a$, where $a$ is the scattering length\cite{olshanii97}.
The parameter $\gamma$ describes the tunnel coupling.

 We are interested in the stability of uniform Josephson oscillations
around the equilibrium configuration where the two condensates have the 
same phase and equal longitudinal density.
 In the sections 2-4, we consider a homogeneous configuration where $U(z)=0$.
 In the sections 2 and 3, we calculate 
the linearized evolution of modes of non zero longitudinal momentum 
in the presence of uniform Josephson oscillations.  
 In the  section 2, we give results of a  calculation valid 
both in the Josephson and in the Rabi regime.
In section 3, 
we  show that, in the Josephson regime,  the system is 
well described by a modified Sine-Gordon equation. 
For small amplitude oscillations, 
we derive scaling laws for the instability time constant and the 
wave vectors of the growing modes.
 In section 4, we go beyond the previous linearized approaches and present
numerical results. We observe damped  oscillations of the uniform Josephson
mode amplitude.
Such oscillations are 
similar to the  Fermi-Pasta-Ulam 
recurrence behavior\cite{Yuen78,Infeld81}.
 In the last section (5), we present numerical calculations in the case of 
a harmonic longitudinal confinement. We show that Josephson oscillations
are stable for a sufficiently strong confinement and we give 
an approximate condition of stability.

\section{Numerical linearized calculation}
\label{sec.deltapsi}
\begin{figure}
\includegraphics{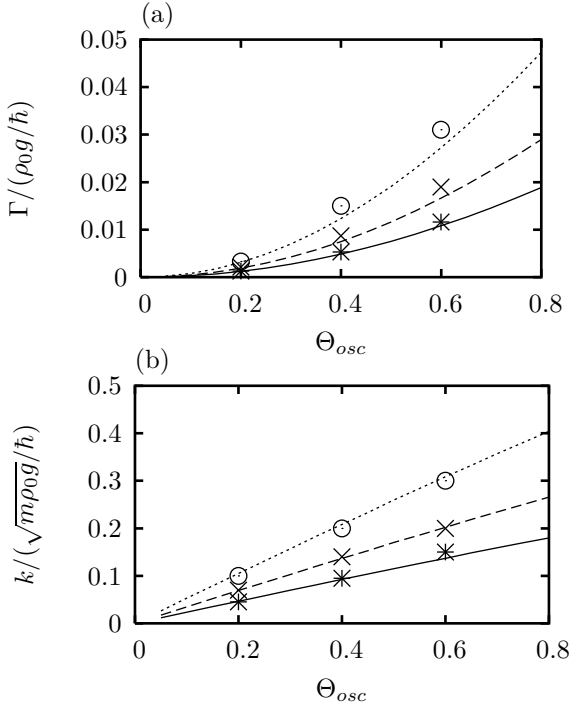}
\caption{\it 
Maximum instability rate of excitations (a) and 
maximum wave vector $k$ of unstable modes (b)
as a function of  the amplitude  of 
the relative phase oscillations for $\gamma=0.05\rho_0 g$ (stars and
solid line) $\gamma=0.1\rho_0 g$ (crosses and dashed line) and
$\gamma=0.2\rho_0 g$ (circles and dotted line). The points are the results
of the linearized numerical calculations presented in section \ref{sec.deltapsi}
and are given with a precision of 10\%.
The continuous lines are given by diagonalising the four by four matrix 
as presented in section \ref{sec.SN}.
}
\label{fig.unstabilities}
\end{figure}

 To investigate whether Josephson oscillations are unstable with 
respect to longitudinal excitations, 
we use a linearized calculation around the time-dependent solution 
corresponding to uniform Josephson oscillations. 
 Writing 
$\psi_1=\varphi_1+\delta\psi_1$ and
$\psi_2=\varphi_2+\delta\psi_2$ with uniform ($z$-independent) 
$\varphi_{1,2}$, Eq.\ref{ham} gives to zeroth order the 
coupled Gross-Pitaevski equations
\begin{equation}
\begin{array}{l}
i\hbar \displaystyle\frac{d}{dt}\varphi_1=g|\varphi_1|^2\varphi_1 -\gamma \varphi_2
+(\gamma -\rho_0 g)\varphi_1\\
i\hbar \displaystyle\frac{d}{dt}\varphi_2=g|\varphi_2|^2\varphi_2 -\gamma \varphi_1
+(\gamma -\rho_0 g)\varphi_2\\
\end{array}.
\label{eq.meanfieldpsi}
\end{equation}
We shifted the zero of energy by adding 
to the Hamiltonian a ``chemical potential'' term 
$\gamma -\rho_0 g$, where 
$\rho_0$ is the density of each condensate at equilibrium.
We recover here the well known results 
established for a two modes model\cite{Legett2001,Sols1999,Smerzi97_Josephson}.
More precisely, writing $\varphi_1=\sqrt{N_1/L}\,\,e^{i\theta_1}$ and
$\varphi_2=\sqrt{(N-N_1)/L}\,\,e^{i\theta_2}$ where $L$ is the 
size of the system, Eq.\ref{eq.meanfieldpsi} implies 
that the conjugate variables $\theta_1-\theta_2$ 
and $k=(N_1-N_2)/2$ evolve according to the non rigid 
pendulum Hamiltonian 
$H_p=E_Ck^2/2+E_J\sqrt{1-4k^2/N^2}\cos(\theta_1-\theta_2)$ where
the charge energy is $E_c=2g/L$ and the Josephson energy 
is $E_J=\gamma N$.
We consider oscillations of $\theta_1-\theta_2$ around 0 of amplitude 
$\Theta_{osc}$. Let us now consider the evolution of excitations
around those uniform oscillations.
To first order in $\delta\psi_{1,2}$, Eq.\ref{ham} yields the 
coupled Bogoliubov equations
\begin{equation} 
i\hbar\frac{d}{dt} 
\left (  
\begin{array}{c}
\delta\psi_1\\ 
-\delta\psi_1^+\\
\delta\psi_2\\
-\delta\psi_2^+\\
\end{array}
\right )=
\left ( \begin{array}{cc}{\cal L}_1& {\cal C}\\{\cal C}&{\cal L}_2 \end{array}\right )
\left ( 
\begin{array}{c}
\delta\psi_1\\
-\delta\psi_1^+\\
\delta\psi_2\\
-\delta\psi_2^+\\
\end{array}
\right )
\label{eq.evoldeltapsi}
\end{equation}
where, for $i=1,2$,
\begin{equation}
\small
{\cal L}_i=\left ( 
\begin{array}{cc}
-\frac{\hbar^2}{2m}
\frac{\partial^2}{\partial z^2}+2g|\varphi_i|^2-\rho_0g+\gamma
& -g\varphi_i^2\\
g{\varphi_i^*}^2 &
\frac{\hbar^2}{2m}
\frac{\partial^2}{\partial z^2}-2g|\varphi_i|^2+\rho_0g-\gamma
\end{array}
\right )
\end{equation}
and the coupling term is 
\begin{equation}
{\cal C}=\left (
\begin{array}{cc}
-\gamma &0\\0&\gamma
\end{array}\right ).
\end{equation}
 Instabilities arise if there exist modes growing exponentially in time
under Eq.\ref{eq.evoldeltapsi}.
The evolution matrix is invariant under translation so that 
we can study independently  
plane waves modes $e^{ikz}(u_1,v_1,u_2,v_2)$,
the second derivatives in ${\cal L}_1$ and ${\cal L}_2$ being replaced by $-k^2$.
 Note that  the evolution of 
excitations depends only on the four parameters $k$, $\rho_0 g$, $\gamma$ 
and $\Theta_{osc}$.
 For a given $k$ component, we numerically evolve  equations   
\ref{eq.meanfieldpsi} and \ref{eq.evoldeltapsi}. 
\begin{figure}
\includegraphics{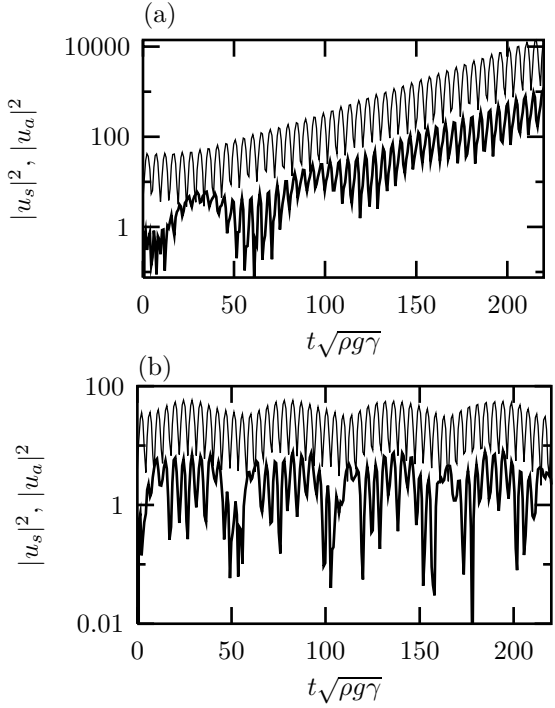}
\caption{\it 
Evolution of the  square amplitude of the symmetric (fat lines) and 
antisymmetric (thin lines) excitations of wave vector 
$k=0.1 \sqrt{m \rho_0 g}/\hbar$ (a) and 
$k=0.3 \sqrt{m \rho_0 g}/\hbar$ (b).
Those graphs are computed for $\gamma=0.1 \rho_0 g$ and a 
uniform Josephson oscillation 
amplitude $\Theta_{osc}=0.6$.
}
\label{fig.evoldeltapsi}
\end{figure}
Fig.\ref{fig.evoldeltapsi} gives the evolution of the square amplitude of the
symmetric mode
$|u_s|^2=|u_1+u_2|^2$ and of the antisymmetric mode 
$|u_a|^2=|u_1-u_2|^2$ for two different
$k$ vectors, for $\gamma=0.1\rho_0 g$ and for $\Theta_{osc}=0.6$. 
For these calculations,
we choose the initial condition as $(u_1,v_1,u_2,v_2)=(1,-1,-1,1)$.
 In the two cases, we observe a fast oscillation at a frequency close to
the frequency of the antisymmetric mode 
$\sqrt{(2\rho_0 g+2\gamma+\hbar^2k^2/2m)(2\gamma+\hbar^2k^2/2m)}$ 
and a slower oscillation at a frequency close to that of the symmetric mode 
$\sqrt{(2\rho_0 g+\hbar^2k^2/2m)\hbar^2k^2/2m}$\cite{Nicketmoi}.
 On top of this, we observe, for $k=0.1$, an exponential 
growth $e^{2\Gamma t}$ of $|u_1+u_2|^2$ and $|u_1-u_2|^2$ , signature
of an unstability. 
 We find that, for given $\rho_0 g$ and $\Theta_{osc}$, 
the instability domain in $k$ is $[0,k_{max}]$.
 Fig.\ref{fig.unstabilities} gives 
the maximum  growth rate $\Gamma$ and the maximum unstable wave vector
$k_{max}$. 

\section{Calculation in the Josephson limit} 
\label{sec.SN}
In this section, we focus on the Josephson regime where $\gamma \ll
\rho_0 g$\cite{notesurFock}.
 In this regime the amplitude 
of oscillations
in the relative density $\delta\rho$ remains small 
compared to the mean density and 
one can assume $\rho_1=\rho_2$ in the Josephson energy term of the
Hamiltonian.
 Furthermore, we restrict ourselves to long wavelength 
excitations described by phonons and 
we neglect anharmonicity of phonons.
Then, the Hamiltonian reduces to
\begin{equation}
H_J=H_s+H_{\scriptsize SG}+H_c,
\label{eq.HamiltonianSG}
\end{equation}
where, writing $\psi_1=\sqrt{\rho_1}e^{i\theta_1}$,
$\psi_2=\sqrt{\rho_2}e^{i\theta_2}$,
$\theta_a=\theta_1-\theta_2$, $\theta_s=\theta_1+\theta_2$, 
$\rho_a=(\rho_1-\rho_2)/2$ and $\rho_s+\rho_0=(\rho_1+\rho_2)/2$,
\begin{equation}
H_s=\int\left ( \frac{\hbar^2\rho_0}{4m}
\left (\frac{\partial \theta_s}{\partial z}\right )^2+g\rho_s^2 \right )dz
\end{equation}
describes the symmetric phonons,
\begin{equation}
H_{\scriptsize SG}=\int \left ( \frac{\hbar^2\rho_0}{4m} 
\left (\frac{\partial \theta_a}{\partial z}\right )^2 +g\rho_a^2
-2\gamma \rho_0 (\cos(\theta_a)-1) \right )dz
\end{equation}
is the Sine-Gordon Hamiltonian 
and
\begin{equation}
H_c=-2\gamma\int\rho_s(\cos(\theta_a)-1)dz
\end{equation}
is a coupling between the symmetric and antisymmetric modes.
The Sine-Gordon Hamiltonian has  already been introduced in
the physics of elongated supraconducting Josephson 
junction\cite{LivreJosephsonsupra}. 
In those
systems, symmetric modes would have a very large charge and magnetic energy and do
not contribute.
 The Sine-Gordon model has been extensively studied\cite{InstabiliteBenjamin}. 
In particular, it has been shown that, for a Sine-Gordon 
Hamiltonian,  oscillations of well defined 
momentum (in particular $k=0$) present Benjamin-Feir 
instabilities\cite{InstabiliteBenjamin}. 
 Our system is not described by the Sine-Gordon Hamiltonian because
of the presence of $H_c$.
In the following, we derive results about
stability of our modified Sine-Gordon system. As we will see later, we recover
results close to that obtained for the Sine-Gordon model.

\begin{figure}
\includegraphics{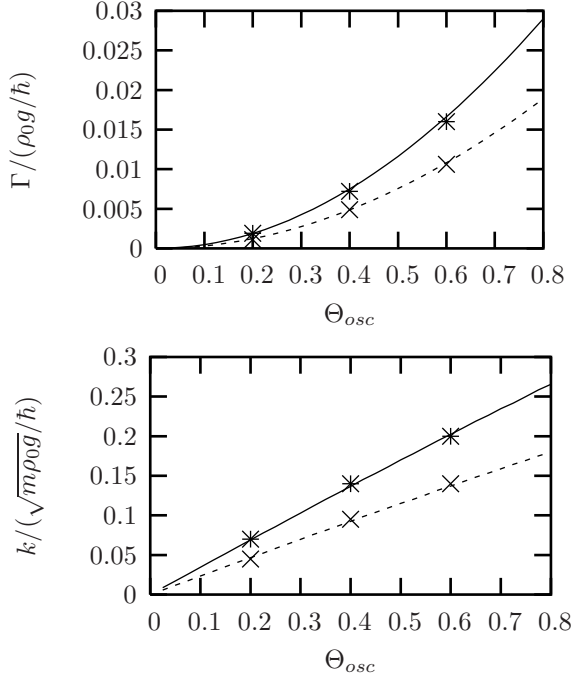}
\caption{\it Comparison between numerical evolution of
  Eqs.\ref{eq.evolthetaosc} and \ref{eq.evoldeltarho}
(points) and the results obtained by diagonalising the 4 by 4 matrix of the 
Floquet representation (lines).
Parameters are $\gamma=0.1\times \rho_0 g$ (stars and continuous lines)
and $\gamma=0.05\times \rho_0 g$ (crosses and dashed line).
}
\label{fig.limiteJosephson}
\end{figure}

 The Josephson oscillations correspond to oscillations where $\rho_a=\rho_{osc}$ and 
$\theta_a=\theta_{osc}$ are independent of $z$. They are given by
\renewcommand{\arraystretch}{2}
\begin{equation}
\left \{
\begin{array}{l}
\displaystyle
\frac{\partial \rho_{osc}}{\partial t} =2\gamma\rho_0\sin(\theta_{osc})/\hbar\\
\displaystyle
\frac{\partial \theta_{osc}}{\partial t} =-2g\rho_{osc}/\hbar\\
\end{array}
\right .
\label{eq.evolthetaosc}
\end{equation}
They  also induce an oscillation $\theta_{osc}^{(s)}$ of $\theta_s$ given by
\begin{equation}
\frac{\partial \theta_{osc}^{(s)}}{\partial t}=-2\gamma
\left (\cos(\theta_{osc}) -1 \right )/\hbar .
\end{equation}
\renewcommand{\arraystretch}{1}
To investigate whether some non vanishing $k$ modes are unstable in 
presence of a Josephson oscillation, we linearize, as in the previous
section,
 the equation of motion
derived from Eq.\ref{eq.HamiltonianSG} around the solution $\rho_{osc}$,
$\theta_{osc}$. Because of translational invariance, we can study independently
the evolution of modes of well defined longitudinal wave vector $k$. 
 Writing $\rho_1=\rho_0+\rho_{osc}+(\delta\rho_a+\delta\rho_s)e^{ikz}$,
$\rho_2=\rho_0-\rho_{osc}+(-\delta\rho_a+\delta\rho_s)e^{ikz}$, 
$\theta_1=(\theta_{osc}^{(s)}+\theta_{osc}+
(\delta\theta_s+\delta\theta_a)e^{ikz})/2$,
and $\theta_2=(\theta_{osc}^{(s)}-\theta_{osc}+
(\delta\theta_s-\delta\theta_a)e^{ikz})/2$,
we find the evolution equation
\begin{equation}
\small
\begin{array}{l}
\hbar\displaystyle\frac{d}{dt}
\left (
\begin{array}{l}
\delta\rho_a/\rho_0\\
\delta\theta_a\\
\delta\rho_s/\rho_0\\
\delta\theta_s
\end{array}
\right )=\\
\left (
\begin{array}{cccc}
0&-\frac{\hbar^2k^2}{2m}+2\gamma\cos(\theta_{osc})&
2\gamma\sin(\theta_{osc})&0\\
-2\rho_0 g&0&0&0\\
0&0&0&-\frac{\hbar^2k^2}{2m}\\
0&2\gamma\sin(\theta_{osc})&-2\rho_0 g&0
\end{array}
\right )
\left (
\begin{array}{l}
\delta\rho_a/\rho_0\\
\delta\theta_a\\
\delta\rho_s/\rho_0\\
\delta\theta_s
\end{array}
\right ) .
\end{array}
\label{eq.evoldeltarho}
\end{equation}
We solved numerically Eq.\ref{eq.evolthetaosc} and Eq.\ref{eq.evoldeltarho}
and we find that modes of low $k$ wave vectors are unstable. 
Fig.\ref{fig.limiteJosephson} gives the instability rate and the maximum 
$k$ wave vector of unstable modes.
 Those results agree within 10\% to the more general results of the previous 
section as long as $\gamma/\rho_0 g< 0.2$ and the oscillation amplitude 
fulfills $\Theta_{osc}<0.6$.

To get more insight into the physics involved and to obtain 
scaling laws for the instability rate and the instability range in $k$, we
will perform several approximations. 
 The evolution matrix $M$ of Eq.\ref{eq.evoldeltarho} is periodic 
in time with a period $\omega_J$. 
We can thus use a Floquet analysis\cite{Floquet} and look for solutions of  
Eq.\ref{eq.evoldeltarho}
in the form 
\begin{equation}
e^{i\nu t}\sum_{n=-\infty}^{+\infty}  e^{in\omega_J t}c_n=
e^{i\nu t}\sum_{n=-\infty}^{+\infty}  e^{in\omega_J t}\left (
\begin{array}{l}
{c_1}_n\\
{c_2}_n\\
{c_3}_n\\
{c_4}_n
\end{array}
\right ).
\end{equation} 
 Expanding Eq.\ref{eq.evoldeltarho} for each Fourier component, we find 
\begin{equation}
\nu c_n=-\omega_J n c_n -i M_0 c_n -i \sum_m M_m c_{n-m},
\label{eq.Floqueteq}
\end{equation} 
where the time independent matrices $M_n$ are the Fourier components
\begin{equation}
M_m=\frac{\omega_J}{2\pi\hbar}\int_0^{\frac{2\pi}{\omega_J}} e^{-im\omega_J t} M(t)dt
\end{equation}
Thus, solutions of Eq.\ref{eq.evoldeltarho} are found as eigenvalues of the linear
set of equations (\ref{eq.Floqueteq}). The mode is unstable if there exists an 
eigenvalue of non vanishing real part and its growth rate is the real part of
the eigenvalue. 

\begin{figure}
\includegraphics{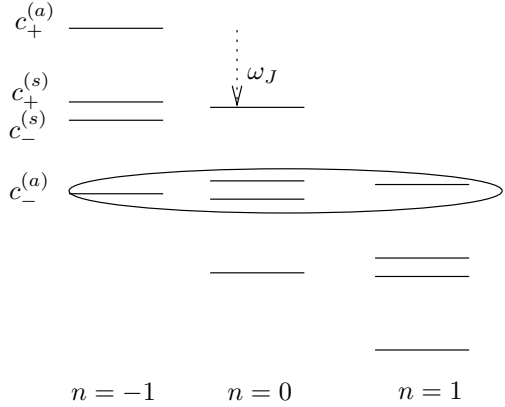}
\caption{\it Floquet representation of the equation Eq.\ref{eq.evoldeltarho}.
The ellipse surrounds the four states that are considered in the calculation
of instability rates.} 
\label{fig.Floquet}
\end{figure}

 For $\Theta_{osc}=0$, only the dc component $M_0$ is not vanishing and
its eigenvalues are 
$\pm \omega_a=\pm i\sqrt{2\rho_0 g(2\gamma+\hbar k^2/2m)}$ 
and 
$\pm \omega_s=\pm i\hbar k\sqrt{\rho_0 g/m}$ corresponding, 
for each Fourier component $n$,
 to the symmetric 
modes ${c^{(s)}_{\pm}}_n$ and antisymmetric modes ${c^{(a)}_{\pm}}_n$.
The four states ${c^{(a)}_{-}}_{-1}$,
${c^{(s)}_{-}}_{0}$, ${c^{(s)}_{+}}_{0}$ and ${c^{(a)}_{+}}_{1}$
form a subspace almost degenerate in energy and of energy far away from 
the other states as depicted Fig.\ref{fig.Floquet}. Thus, we will restrict 
ourselves to those states in the following.
 In the limit of oscillations of 
small amplitude $\Theta_{osc}$, the matrix elements of $M$
can be expanded to second order in $\theta_{osc}$. 
Furthermore, the oscillations are well described by
$\theta_{osc}=\Theta_{osc}\cos(\omega_J t)$, where    
$\omega_J=2\sqrt{\gamma\rho_0g}(1-\Theta_{osc}^2/16)/\hbar$.
 We then find that, in the 4 dimensional subspace spanned by
(${c^{(a)}_{-}}_{-1}$,
${c^{(s)}_{-}}_{0}$, ${c^{(s)}_{+}}_{0}$,${c^{(a)}_{+}}_{1}$),
the eigenvalue $\nu$ of Eq.\ref{eq.Floqueteq} are the eigenvalues 
of the four by four matrix
\renewcommand{\arraycolsep}{0.15cm}
\renewcommand{\arraystretch}{1.5}
\begin{equation}
{\cal M}=\left ( \begin{array}{cccc}
i(-\omega_a+\omega_J+\gamma\Theta_{osc}^2 f_a^2/4)&
-i\gamma\Theta_{osc}f_a/(2 f_s) &\gamma\Theta_{osc}f_a/(2 f_s)&
-\gamma \Theta_{osc}^2 f_a^2/8\\
i\gamma\Theta_{osc}f_a/f_s/2&
i\omega_s& 0 &-\gamma \Theta_{osc}f_a/(2 f_s)\\
_\gamma\Theta_{osc}f_a/(2 f_s)&
0&-i\omega_s&-i\gamma\Theta_{osc}f_a/(2 f_s)\\
-\gamma \Theta_{osc}^2f_a^2/8& \gamma\Theta_{osc}f_a/(2f_s)&
i \gamma\Theta_{osc}f_a/(2f_s)&
i(\omega_a-\omega_J-\gamma\Theta_{osc}^2 f_a^2/4)\\
\end{array}
\right )
\label{eq.M}
\end{equation}
where $f_a=(2\rho_0 g/(\hbar^2k^2/2m+2\gamma))^{(1/4)}$ and
$f_s=(4m\rho_0 g/\hbar^2 k^2)^{(1/4)}$.

We  numerically diagonalise this matrix and find the instability rate
as the largest real part of the eigenvalues. 
 For a given oscillation amplitude $\Theta_{osc}$, scanning the wave
vector $k$, we find the largest instability rate and the
maximum wave vector of unstable modes. 
Fig.\ref{fig.limiteJosephson} compare those results 
with the values obtained by integration of 
Eq.\ref{eq.evoldeltarho}.
 We find 
a very good agreement in the range $\theta<0.6$ and $\gamma/\rho_0
g<0.1$. Finally, in Fig.\ref{fig.unstabilities}, 
we compare the instability rate and the
maximum unstable wave vector found with this simplified Floquet analysis
with the more general results of section 1. We find a very good
agreement as long as $\gamma/\rho_0 g<0.2$ and $\Theta_{osc}<0.6$.

\begin{figure}
\includegraphics{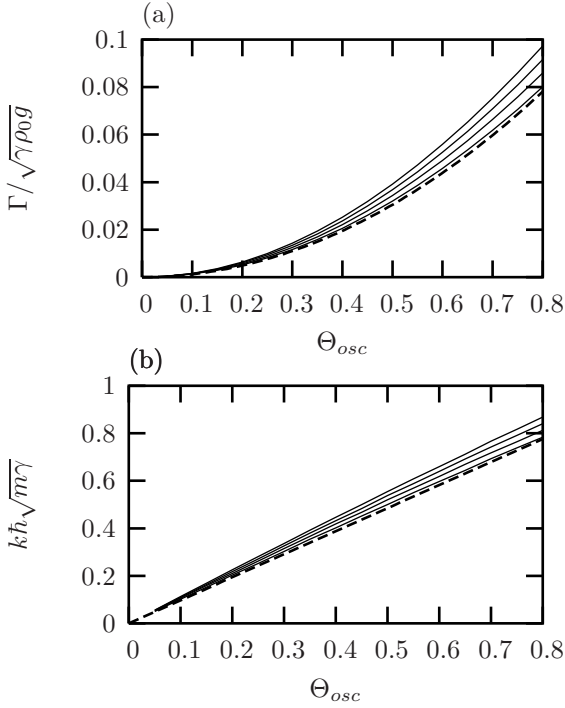}
\caption{ Maximum instability rate  normalized to the Josephson oscillation frequency 
(a) and maximum wave vector of unstable modes  normalized to 
$\sqrt{m\gamma}/\hbar$ (b)
 as a function of the oscillations
amplitude $\Theta_{osc}$
for different ratios $\gamma/\rho_0 g$ (from lower curves to upper curves :
0.02,0.06,0.1,0.14). 
Fat dashed lines are the scaling laws Eq.\ref{eq.gammainstscaling} and
\ref{eq.kinstscaling}. Thin continuous lines are found by 
diagonalising the matrix of Eq.\ref{eq.M}.
}
\label{fig.scalinglaws}
\end{figure}

 If we restrict ourselves to terms linear in $\Theta_{osc}$, then 
the only effect of the Josephson oscillations is to introduce
a coupling between the symmetric and antisymmetric mode.
We checked that  this coupling alone does
not introduce any instability. 
 Thus instability is due to the quadratic terms. Those terms contain a modulation
at $2\omega_J$. This modulation corresponds to the modulation of the 
frequency of the antisymmetric mode
\begin{equation}
\omega_a^2=2\rho_0g(\hbar^2 k^2/2m+2\gamma-2\gamma\Theta_{osc}^2/4)+
\gamma\rho_0g\Theta_{osc}^2\cos(2\omega_J t).
\end{equation} 
  This parametric oscillation leads to instability for 
$k\in [0,\Theta_{osc}\sqrt{m\gamma/2}/\hbar]$ and the 
instability time constant at resonance is 
$\Gamma=\Theta_{osc}^2\frac{\sqrt{\gamma\rho_0g}}{8\hbar}$.
We recover here the well known results 
of Benjamin-Feir instability derived for example in \cite{InstabiliteBenjamin} 
using the multiple-scale perturbation technique.
In our case,
the  coupling to the symmetric mode will modify those values.
However, for small values of $\gamma$, the qualitative behavior is
unchanged.
Indeed, as
seen in Fig.\ref{fig.scalinglaws}, as long as $\gamma<0.05\rho_0 g$
and within a precision of 10\%,
the instability rate $\Gamma$ 
scales as
\begin{equation}
\Gamma=0.122(1)\Theta_{osc}^2\sqrt{\gamma\rho_0 g}/\hbar
\label{eq.gammainstscaling}
\end{equation}
 and the  maximum wave vector of unstable modes as 
\begin{equation}
k_{max}=0.97(1) \frac{\sqrt{m\gamma}}{\hbar}\Theta_{osc}.
\label{eq.kinstscaling}
\end{equation}
For larger $\gamma$, the $\Gamma$ and $k_{max}$ are higher than 
those lows as seen in Fig.\ref{fig.scalinglaws}.

\section{Beyond the linearisation}
\begin{figure}
\begin{center}
\includegraphics{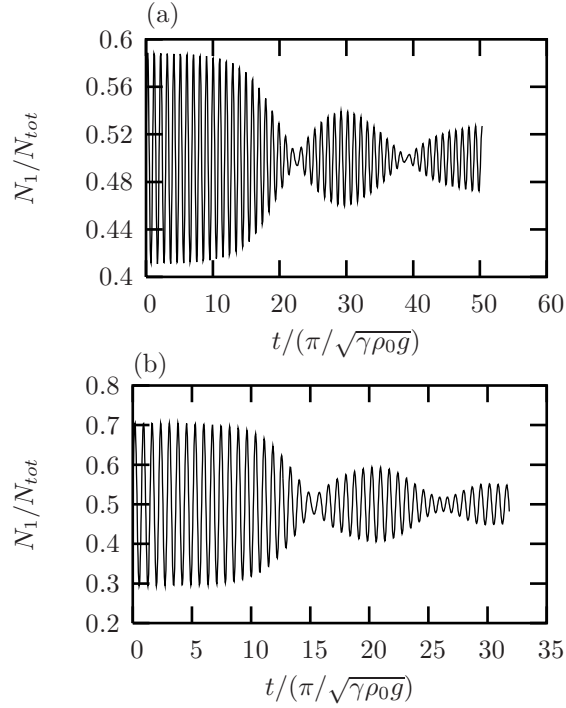}
\end{center}
\caption{\it Evolution of the number of atoms in the condensate 1, 
normalized 
to the total number of atoms,
 as a 
function of time.
The initial state corresponds to a phase difference between the condensate
$\Theta_{osc}=0.6$ superimposed on
phase and density fluctuations corresponding to a 
thermal equilibrium at temperature 
$k_B T=0.1\rho_0g$.
For this calculation, $\gamma=0.1\rho_0 g$ (a) and $\gamma=\rho_0 g$ (b).}
\label{fig.n1moy}
\end{figure}

 The two previous sections give a linearized analysis of the evolution 
of perturbations. They show that the presence of uniform Josephson oscillations 
produces instabilities of modes of non vanishing momentum. 
 The energy in these mode grows and consequently, the energy of the 
uniform Josephson mode decreases and one expects a decrease 
of the uniform Josephson oscillations amplitude. 
 Such a decrease is beyond the previous linearized analysis and 
we perform full numerical calculation of the evolution of the mean fields
$\psi_1(z,t)$ and $\psi_2(z,t)$.
 The evolution equations derived from Eq.\ref{ham} are
\renewcommand{\arraystretch}{2}
\begin{equation}
\left \{
\begin{array}{l}
i\hbar
\displaystyle\frac{d}{dt}\psi_1=-\displaystyle\frac{\hbar^2}{2m}
\displaystyle\frac{d^2\psi_1}{dz^2}+g|\psi_1|^2\psi_1
-\gamma\psi_2\\
i\hbar\displaystyle\frac{d}{dt}\psi_2=-\displaystyle\frac{\hbar^2}{2m}\displaystyle\frac{d^2\psi_2}{dz^2}+g|\psi_2|^2\psi_2
-\gamma\psi_1\\
\end{array}
\right .
\label{eq.coupledGP}
\end{equation}
Fig.\ref{fig.n1moy} gives
 the evolution of the total number of atoms in 
the condensate 1, $N_1=\int |\psi_1|^2$, for initial amplitude 
$\Theta_{osc}=0.6$ and for different values of $\gamma/(\rho_0 g)$.
For these calculations, the initial state consists in 
a z-independent phase difference $\Theta_{osc}$
between $\psi_1$ and $\psi_2$ superposed on thermal fluctuations 
of the density and phase of the two condensates corresponding to 
a temperature $k_BT=\rho_0 g/10$. 
 We observe that the amplitude of the Uniform Josephson  Oscillations 
presents damped oscillations. 
For $\gamma\ll \rho_0 g$, 
the period of these amplitude oscillations is about 
three times the inverse of the 
instability rate of Eq.\ref{eq.gammainstscaling}. The ratio 
between the Josephson frequency and the 
frequency of these amplitude oscillations 
is about 20 and is almost
independent on the ratio between $\gamma$ and $\rho_0 g$ as long as 
$\gamma <\rho_0 g$.  
 For larger $\gamma$, this  ratio increases and more 
Josephson oscillations are seen in a period of the amplitude modulation.
 Such amplitude oscillations are a reminiscence of the Fermi-Ulam-Pasta
recurrence behavior observed in many
non linear systems with modulational 
instabilities\cite{InstabiliteBenjamin,Yuen78,Infeld81}.
 In particular, this recurrence behavior has been seen in numerical 
evolution of the Sine-Gordon Hamiltonian\cite{Barday91}.
 In our case, we observe an additional damping which 
results probably from the coupling to symmetric modes. 

\begin{figure}
\includegraphics{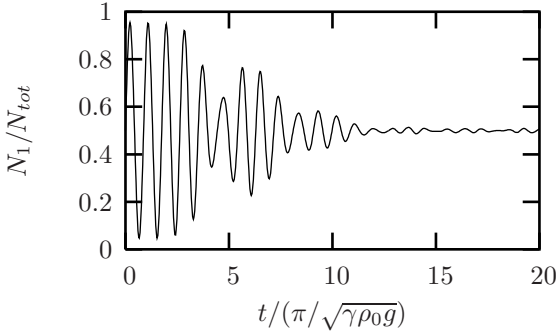}
\caption {\it 
Evolution of the number of atoms in the condensate 1, normalized 
to the total number of atoms, as a function of time
for $\gamma=\rho_0 g$ and an initial phase difference between condensates 
$\Theta_{osc}=\pi/2$. Initial thermal population of excited modes
corresponding to $k_B T=0.1\rho_0g$ is assumed.}
\label{fig.pis2}
\end{figure}

The case of an initial amplitude $\Theta_{osc}=\pi/2$ is of 
particular interest as, in absence of interactions between 
atoms, it corresponds to Rabi oscillations of maximum amplitude.
Fig.\ref{fig.pis2} gives the evolution of $N_1$ for $\gamma=\rho_0 g$
and an initial amplitude $\Theta_{osc}=\pi/2$.

\section{Case of a confined system}
 In the previous sections, we considered large and homogeneous systems.
We found that unstable excited modes are those of low wave vectors. 
In the Josephson limit where $\gamma\ll \rho_0 g$, we derived the 
scaling law Eq.\ref{eq.kinstscaling} for the maximum unstable 
wave vector.
 In a cloud trapped in a box like potential of extension $L$,
the minimum $k$ value
of the excitation modes is $2\pi/L$. 
Thus, if 
\begin{equation}
L < \frac{2\pi\hbar}{1.0\sqrt{m\gamma}\Theta_{osc}},
\label{eq.sizemax}
\end{equation}
the minimum wave vector of excited modes is
larger than 
the maximum unstable $k$ value Eq.\ref{eq.kinstscaling} and
the system is stable.  This condition can be understood in
a different way~:
the energy of the lowest longitudinal mode is $\hbar 2\pi\sqrt{\rho_0 g}/(m L)$ (
here we assume $L\gg \hbar/\sqrt{m\rho_0 g}$). Thus, we find that the system 
is stable provided that the energy of the lowest excited mode satisfies
$E_{exc} > 0.52 \omega_J \Theta_{osc}$
where $\omega_J=2\sqrt{\gamma\rho_0 g}/\hbar$ is the Josephson frequency.

\begin{figure}
\includegraphics{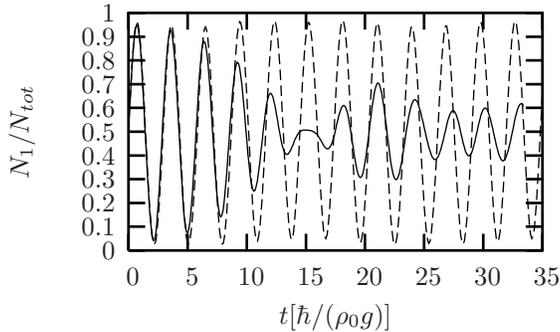}
\caption{Josephson oscillations of clouds trapped in a harmonic potential
of frequency $\omega=0.1 \rho_0 g/\hbar$ (solid line) and 
$\omega=\rho_0 g/\hbar$ (dashed line), where $\rho_0$ is the peak 
linear density in each condensate. The initial phase difference 
between the condensates is $\pi/2$ 
and the tunnel coupling is $\gamma=\rho_0 g$.
$N_1$ is the number of atoms in the condensate 1 and $N_{tot}$ the 
total number of atoms.} 
\label{fig.piege}
\end{figure}

 An  approximate condition of stability of Josephson oscillations 
in the case of a cloud trapped in a  harmonic longitudinal potential
of frequency $\omega$ is found as follows.
 The  size of cloud, described by a Thomas Fermi profile, 
is $L=2\mu/(m\omega^2)$, where $\mu=\rho_0 g$ is the chemical 
potential and $\rho_0$ the peak linear density. Then, from the same
argument as above, one expects to observe stable oscillations
for
\begin{equation}
\omega >\alpha \sqrt{\gamma\rho_0g}\Theta_{osc}=\alpha\Theta_{osc}\omega_J/2
\end{equation}
where $\alpha$ is a numerical factor close to one.
 We performed numerical simulations of the evolution in the 
case of a harmonic potential, adding to both left hand 
sides of Eqs.\ref{eq.coupledGP}
a trapping potential $1/2 m\omega^2 z^2$.
 The initial situation is the Thomas Fermi profile 
superposed on thermal random fluctuations and 
a global phase difference between the condensates 
$\Theta_{osc}=\pi/2$. The tunnel coupling is $\gamma=\rho_0 g$.
The resulting Josephson oscillations are shown in
Fig.\ref{fig.piege} for $\omega=\rho_0 g/\hbar$ and
$\omega=0.1\rho_0 g/\hbar$.
We observe that for  $\omega=\rho_0 g/\hbar$, Josephson 
oscillations are stable whereas, for  $\omega=0.1 \rho_0 g/\hbar$,
oscillations are unstable.

\section{Conclusion and prospects}
 We have shown that Josephson oscillations
of two coupled elongated condensates are unstable with respect to 
excitations of longitudinal modes. The unstable modes are
those of small wave vectors.
 In the Josephson limit where $\gamma\ll \rho_0 g$, we have 
derived the scaling lows Eq.\ref{eq.gammainstscaling} and 
Eq.\ref{eq.kinstscaling} for the instability time constant 
and wave vectors. Since the frequency of Josephson 
oscillations are $2\sqrt{\gamma\rho_0 g}$,  the first equation tells us that  
the number of oscillations that can be observed scales as $\Theta_{osc}^2$
and is independent on $\gamma/\rho_0 g$. This is true as long as 
$\gamma < \rho_0 g$. For larger $\gamma/(\rho_0 g)$, the Josephson 
condition is not fulfilled. Effect of interactions is less pronounced
and more oscillations can be observed.
 Performing numerical calculations beyond the linearized approach,
we have shown that the system presents a recurrence behavior, although
it is damped quickly.
 Finally, we investigated the stability of oscillations 
in finite size systems. 
 Eq.\ref{eq.sizemax} gives the maximum longitudinal size 
of confined condensate that enables the presence of stable
Josephson oscillations. We also considered the case of harmonically
trapped cloud and give an approximate condition on the 
oscillation frequency to have stable Josephson oscillations.

 The results of this paper are not changed drastically for
finite temperature.  
Indeed, although elongated Bose-Einstein condensates present
thermally excited longitudinal phase 
fluctuations\cite{Dettmer2002,Richard2002}, it has 
been shown in \cite{Nicketmoi}
that, because the antisymmetric modes present an energy gap,
thermal fluctuations of the relative phase between elongated coupled  
condensates are strongly suppressed.

 Among the possible extensions of this work, two questions are of 
immediate experimental interest.
 First, the effect of a random longitudinal potential could be
investigated. Indeed, it has been proposed to 
realized elongated coupled condensates
using magnetic trapped formed by micro-fabricated
wires\cite{SchummJosephson}, but,
for such systems, a roughness of the longitudinal potential
has been observed\cite{Zimmermann-fragPRA2002,Hinds-frag2003,Nous-fragm}. 
If the amplitude of the roughness potential 
is smaller than the chemical potential of the condensate, one expects 
to still have a two single elongated condensate. However, 
the roughness of the potential may change significantly the results of 
this paper.
Second, the effect of correlations between atoms may be studied.
Indeed, for large interactions between atoms, correlations 
between atoms become important. More precisely, for 
$\rho_0<mg/\hbar^2$,  a mean 
field approach is wrong and the gas is close to the
Tonks-Girardeau regime\cite{TonksBloch,Tonksomegaosc,TonksBill}.
Such a situation is not described in 
this paper in which a mean field approach has been assumed.
Thus, a new study should be devoted to the physics of coupled 
elongated Tonks gas.

 Dynamical instabilities of the uniform Josephson mode are not the
only effect of non linearities in the system of two coupled elongated
condensates and other interesting phenomena are expected. For instance, 
ref.\cite{JosephsonVortices} shows that Josephson vortices similar to the
solitons of the Sine-Gordon model exist for large enough interaction
energy.

We thank Dimitri Gangardt for 
helpful discussions. 
This work was supported by EU (IST-2001-38863, MRTN-CT-2003-505032),
DGA (03.34.033) and by the French ministery of research (action
concertée
``nanosciences'').


\begin{thebibliography}{10}
\bibitem{notesurFock} We also consider that $\gamma \gg g\rho_0/N^2$, $N$
being the total number of atoms, so that quantum fluctuations 
of the relative phase are negligible.
\end{thebibliography}

\end{document}